\documentclass[aps,11pt,twoside, nofootinbib]{revtex4}

\usepackage{graphicx}
\usepackage{epsfig}
\usepackage{dcolumn}
\usepackage{amsmath,amssymb,amsthm}
\usepackage{bm}
\usepackage{verbatim}
\usepackage{natbib}

\usepackage{color}

\newcommand{\ket}[1]{|#1\rangle}

\newcommand{\ketbra}[2]{| #1 \rangle \langle #2 |}

\begin{document}
\title{ Quantum state of a free spin-$\frac{1}{2}$  particle and
the inextricable dependence of spin and momentum  under Lorentz
 transformations} 
\author{Tiago Debarba} 
\email{debarba@fisica.umfg.br}
\author{Reinaldo O. Vianna}
\email{reinaldo@fisica.ufmg.br}

\affiliation{Departamento de F\'{\i}sica - ICEx - Universidade Federal de Minas Gerais,
Av. Pres.  Ant\^onio Carlos 6627 - Belo Horizonte - MG - Brazil - 31270-901.}

\date{\today}
\begin{abstract}

{
We revise the Dirac equation for a free particle and investigate Lorentz
transformations on spinors. We study how the spin quantization axis changes
under Lorentz transformations and evince the interplay between spin and momentum 
in this context. }

\end{abstract}
\maketitle
\section{Introduction}

Authors like Alsing and Milburn \cite{milburn2002}, or Hacyan \cite{hacyan2001} 
concluded that entanglement is invariant under Lorentz transformations. 
Accordingly, Terashima and Ueda \cite{ueda2002} showed that, under Lorentz
transformations, the perfect anti-correlation of a singlet state is 
recovered if one appropriately corrects the detector's orientation 
according to the Wigner rotation \cite{wigner39}.
Caban and Remienlinski also did an interesting study of EPR correlations
in the quantum field theory context\cite{caban06}.
 On the other hand, Peres {\em et al.} 
\cite{peres2002,peres2004} considered a single free spin-$\frac{1}{2}$
particle, and literally stated that `` the reduced density matrix for its
spin is not covariant under Lorentz transformation. The spin entropy
is not a relativistic scalar and has no invariant meaning". This later
conclusion by Peres {\em et al.} \cite{peres2002} may lead  
a careless reader to  think that  a mere change of reference frame
could create entanglement. 
In Quantum Information, {\em entanglement} is a resource  which allows
for the realization of teleportation \cite{bennett}, quantum cryptography
\cite{crypto}, quantum-enhanced global positioning \cite{gps}, just
to cite a few impressive possible applications.
Though the Wigner rotation acts only on the spin degrees of freedom, it
is not independent of the momentum, and a Lorentz transformation 
cannot be reduced solely to a Wigner rotation, without altering the
momentum. In the relativistic context, spin and momentum are not
independent degrees of freedom. It is not possible a change of 
inertial reference frame, which changes the momentum, i.e. a boost, 
without changing the quantization axis of the spin. 
Among others, this was realized by Gingrich and Adami \cite{adami}, who 
 concluded that ``while spin and momentum entanglement separately
are not Lorentz invariant, the joint entanglement of the wave function is."
As a matter of fact, it is simple to understand. If we insist in facing
spin and momentum as independent degrees of freedom in the relativistic
context, which they are not, a change of inertial reference frame is a
global linear operation acting on spin and momentum. Even when one
chooses a unitary representation for the inhomogeneous Lorentz group, and then
the Wigner rotation comes into play, a Lorentz boost acts as a 
global unitary on spin and momentum, and not as local unitary solely
on spin, or solely on momentum. A way out of this subtleties was proposed
by Bartlett and Terno \cite{bartlett2005} in their ``relativistic
invariant quantum information" approach. 
Notwithstanding one can find in the literature calculations showing
the variation  of spin entanglement solely due to a Lorentz transformation
\cite{contraexemplo}. 

The inextricable dependence of spin
and momentum, in the  relativistic quantum information context, was
discussed by Czachor and Wilzcewski \cite{czachor03}, who considered the 
relativistic version of the Bennett-Brassard \cite{bb84} cryptographic scheme.
Even though, the subject is  controversial, as exemplified by  Czachor's
\cite{comment}  comment
on Peres {\em et al.} work  \cite{peres2002}. Following ideas similar to
Czachor's,  Caban and Rembieli\'nski \cite{caban 2005} developed a 
covariant reduced spin density matrix, with the caveat that  the momentum 
dependent Lorentz transformations could not be represented unitarily
and should act on the Dirac spinors rather than on  {\em spin kets}.
Czachor recently wrote an extensive study of the covariance of quantum 
information \cite{czachor10}, and also concluded that the 
entanglement does not change under Lorentz transformations.

As one can easily realize, the quantum information community is still
trying to come to grips with the proper treatment of entanglement 
in the relativistic context. 
The present work is an attempt to guide the newcomer straightforwardly
 to the conclusion of the covariance of entanglement, using a very simple
but yet rigorous formalism.  The main difference of our approach and 
similar works in the literature is the way we treat the kinematic 
degrees of freedom. We explicitly  identify the single particle Hilbert
 space as labeled by both spin and momentum, calling attention to the
fact that they do not compose by means of a tensor product and, 
therefore,  have no well defined partial trace, in the usual sense, 
but which does not hinder one to take averages on momentum, as done by
Caban and Rembieli\'nski \cite{caban 2005}. The other crucial point
is that we do not single out any particular spin operator, like
Pauli-Lubanski, rather we obtain the correct basis for spin measurements
directly from the form of the operators in the Lorentz algebra.
This is illustrated by considering two observers in relative motion, 
sending particles with well defined momentum to each other and
performing spin measurements, a scenario representative of 
quantum information exchange between an Earth station and
a satellite \cite{satellite}.

The paper is organized as follows. 
We start by reviewing 
the Dirac equation for a free particle. 
We then investigate
how the free spin-$\frac{1}{2}$ particle state transforms under an
inertial reference frame change. In order to do that, we use
the invariance of the Hilbert space internal product \cite{wigner39} and
derive a non-unitary representation of the Lorentz group. Our motivation
to do this is that the choice of representation does not change
the {\em physics}, and thus we can highlight the interdependence of spin
and momentum  in the Lorentz transformation. In the sequence, 
 we verify  that the
purity (or mixedness) of the state is Lorentz invariant, and show
how the spin quantization axis changes with the momentum. Finally we
do some illustrative calculations on how to perform the momentum
dependent  spin measurements on moving frames, and conclude.

\section{ Dirac equation for a free particle}

In this section we recall the arguments that lead to the 
fundamental equation of the relativistic quantum mechanics, the Dirac equation
\cite{dirac,peskin,greiner}, whose solution is a particle with 4-momentum 
$p=(E,\vec{p})$ and spin $1/2$ (or an antiparticle 
with 4-momentum $p=(-E,\vec{p})$).
We first write a  Schr\"odinger like equation
 (natural units are assumed, $\hbar=c=1$):
\begin{equation}
 i\frac{\partial \Psi(x)}{\partial t}=H \Psi(x),
\end{equation}
where the variable $x=(t,\vec{x})$ is the 4-position in the Minkowski space.
The Hamiltonian $H$ is the relativistic energy $H=\sqrt{|\vec{p}|^2+m^2}$,
 where $\vec{p}$ is the momentum vector and $m$ is the mass in rest frame.
 The Dirac equation is obtained  by imposing  linear derivatives on both  space 
and time in Eq.1, which guarantees  Lorentz covariance 
(Lorentz invariance of form), i.e. the physical content of the equation is the
 same in all relativistic inertial frames. 
Therefore, the {\em Dirac Hamiltonian} 
 has the form:
\begin{equation}\label{1}
      H=\vec{\alpha}\cdot\vec{p}+\beta m, 
\end{equation}
and the Dirac equation for a free particle reads:
\begin{equation}
 i\frac{\partial \Psi(x)}{\partial t}=(\vec{\alpha}\cdot\vec{p}+\beta m )\Psi(x).
\end{equation}
As the Dirac Hamiltonian should be consistent with  the relativistic energy,
 i.e. $H^2=(\vec{\alpha}\cdot\vec{p}+\beta m)^2=|\vec{p}|^2+m^2$,
the constants $\alpha_{i}$ and $\beta$ must satisfy
the  following relations:
\begin{align} \label{2}
  &\alpha_{i}\alpha_{j}+\alpha_{j}\alpha_{i}=2\delta_{i,j}, \\
\label{3}
  &\alpha_{i}\beta+\beta\alpha_{i}=0, \\
\label{4}
  &\beta^2=1.
\end{align}
The simplest solution to the above relations is $4\times 4$ traceless unitary
Hermitian matrices \cite{peskin,greiner}.
 Thus the Dirac equation is a $4\times4$ matrix equation.
The solutions of the Dirac equation, known as {\em spinors}, 
 have four components,  like the 4-vectors in Minkowski space, 
but they  do not transform like vectors under Lorentz transformations. 
The {\em spinors} call for a  new representation of the Lorentz transformations.

A possible choice for the $\alpha_{i}$ and $\beta$ matrices,
known as Weyl or Chiral representation (see for example Appendix A.3 in 
\cite{peskin}) is: 
\begin{align}
\alpha_{i}=
\begin{bmatrix}
		 \sigma_{i} & 0 \\
		          0 & -\sigma_{i} \\
\end{bmatrix}
\quad \text{and} \quad
\beta=
\begin{bmatrix}
		0 & I \\
		I & 0 \\
\end{bmatrix},
 \end{align}
where $\sigma_{i}$ are the Pauli matrices and $I$ is the $2\times 2$ identity.
We now define the  \textit{gamma matrices},  $\gamma^{i}=\alpha^{i}\beta$ 
and $\gamma^{0}=\beta$, such that we have a 4-vector of matrices
 $\gamma^{\mu}=(\gamma^{0},\vec{\gamma})$, i.e.:
\begin{equation}
\gamma^{\mu}=
\begin{bmatrix}
		0 & \sigma^{\mu} \\
		\overline{\sigma}^{\mu} & 0 \\
\end{bmatrix}.
\end{equation}
It is convenient to define the Pauli matrix 4-vectors
$\sigma^{\mu}=(I,\vec{\sigma})$ and 
$\overline{\sigma}^{\mu}=(I,-\vec{\sigma})$. 
Finally, the  the Dirac equation in covariant form reads:
\begin{equation}
(\gamma^{\mu}p_{\mu}-m)\Psi(x)=0.
\end{equation}     

For a free particle with 4-momentum $p=(E,\vec{p})$, in a given  reference frame,
 the Dirac equation solution is given in the form:
\begin{equation}\label{9}
\Psi(x)=\exp{(-ip\cdot x)}u(p).
\end{equation}
Substituting this function in the Dirac equation, we obtain for the
spinors $u(p)$:
\begin{equation}\label{5}
(\gamma^{\mu}p_{\mu}-m)u(p)=0.
\end{equation}
For a particle in its rest frame, with 4-momentum  $p_{0}=(m,0,0,0)$,
Eq.\ref{5} reduces to:
\begin{align} 
m
\begin{bmatrix}
		-I & I \\
		I & -I \\
\end{bmatrix}
u(p_{0})=0.
\end{align}
The solutions of Eq.12 are of the form:
\begin{align}
u(p_{0})=N\begin{bmatrix}
	\xi \\ 
	\xi 
\end{bmatrix},
\end{align}
where $N$ is a normalization factor. 
 Since the Dirac equation is a $4\times 4$ matrix equation, the  $\xi$
 must be 2-component vectors and  form an  orthonormal basis 
in the Hilbert space. Therefore we write:
\begin{align}
\xi^{0}=
\begin{bmatrix}
	1 \\ 
	0
\end{bmatrix}
\quad \text{and} \quad
\xi^{1}=
\begin{bmatrix}
	0 \\ 
	1
\end{bmatrix}.
\end{align}
These vectors label a new degree of freedom, that we had not taken into
 account, the particle's spin. Therefore, $u(p_{0})$  needs one more label,
the spin orientation ($\alpha$): 
\begin{align}\label{6}
  u(p_{0},\alpha)=N\begin{bmatrix}
	\xi^{\alpha} \\ 
	\xi^{\alpha} 
\end{bmatrix}.
\end{align}
As we are in the particle's rest frame, the chosen basis for the vectors
 $\xi^{\alpha}$ are usually the eigenstates of the  $\sigma_{z}$ Pauli matrix,
 so the spin quantization axis is in $z$ direction and we have the spin
 orientations $+1/2$ when $\alpha=0$ and $-1/2$ when $\alpha=1$.

If we now consider a free antiparticle with negative energy,
\begin{equation}
\Psi(x)=\exp{(ip\cdot x)}v(p,\alpha),
\end{equation}
the Dirac equation for the spinors $v(p,\alpha)$ reads
\begin{equation}
(\gamma^{\mu}p_{\mu}+m)v(p,\alpha)=0,
\end{equation}
and the solution in the   rest frame is:
\begin{align}
v(p_{0},\alpha)=M
\begin{bmatrix}
		\eta^{\alpha} \\
		-\eta^{\alpha} \\
\end{bmatrix},
\end{align}
with $\eta^{0}=\begin{bmatrix}
	0 \\ 
	1
\end{bmatrix}$  and $\eta^{1}=\begin{bmatrix}
	1 \\ 
	0
\end{bmatrix}$.

\section{Lorentz Transformations on Dirac Spinors}
In this section, we obtain the representation of the Lorentz transformations
that act on the spinors. Let  $S^{\mu\nu}$ and $S^{\rho\sigma}$  be elements
of the Lorentz algebra, satisfying the following commutation relation 
\cite{peskin,greiner}:
\begin{equation}\label{101}
[S^{\mu\nu},S^{\rho\sigma}]=i(g^{\nu\rho}S^{\mu\sigma}-g^{\mu\rho}S^{\nu\sigma}-g^{\nu\sigma}S^{\mu\rho}+g^{\mu\sigma}S^{\nu\rho}).
\end{equation}
The metric of Minkowski space is $g^{\mu\nu}=diag(1,-1,-1,-1)$,
which can be obtained from the anti-commutation relation of  the gamma matrices
 $g^{\mu\nu}=\frac{1}{2}\{\gamma^{\mu},\gamma^{\nu}\}$.
On the other hand, the $S^{\mu\nu}$  can be obtained from the commutation
 relation of the gamma matrices:
\begin{equation}
S^{\mu\nu}=\frac{i}{4}[\gamma^{\mu},\gamma^{\nu}].
\end{equation}
This algebra has six elements, namely  three  for the boosts (translations),
and other three for spin rotations.
 $S^{0k}$ is the Lorentz boost generator in the direction $k=1,2,3$, 
 while $S^{ij}$ is the angular momentum operator of the spin,
 responsible for its  rotation in the plane $i,j=1,2,3$.
Note that as consequence of the tensorial character of the spin
  operator $S^{ij}$ \cite{peskin}, 
spin observables intrinsically have
tensorial character in   the  context of relativistic quantum mechanics.

Now writing  explicitly  the Lorentz operators:
\begin{align}\label{12}
S^{0k}=\frac{-i}{2}
\begin{bmatrix}
		\sigma_{k} & 0 \\
		0 & -\sigma_{k} \\ 
\end{bmatrix}
\quad \text{and} \quad
S^{ij}=\frac{1}{2}\epsilon_{ijk}
\begin{bmatrix}
		\sigma_{k} & 0 \\
		0 & \sigma_{k} \\ 
\end{bmatrix},
\end{align}
we obtain the Lorentz group as an exponential function of the
 Lorentz algebra. If $D(\omega)$ is an element of the group,
 and  $\omega$ is the transformation parameter, we have: 
\begin{equation}\label{15}
D(\omega)=\exp{\Big(\frac{-i}{2}\omega_{\mu\nu}S^{\mu\nu}\Big)}.
\end{equation} 
The boost representation will be 
$D(\omega_{k})=\exp{(\frac{-i}{2}\omega_{0k}S^{0k})}$, 
where $\omega_{0k}=2\eta_{k}$, being $\eta_{k}$ the rapidity of
 the particle in direction $k$. The rapidity can be defined as a
 function of the relative velocity, $\tanh{\eta}=\beta$. 
Therefore, we can write the boosts having the rapidity as a 
parameter of the transformation: $D(\eta)=\exp{(-iS^{0k}\eta_{k})}$.
 In the case that the particle has a 4-momentum $p=(E,\vec{p})$, 
the rapidity is $\cosh{\eta}=E/m$ and $\sinh{\eta}=p/m$, this results in
 the Lorentz boost:
\begin{align} \label{15b}
D(\eta)=\frac{1}{\sqrt{2m(E+m)}}
\begin{bmatrix}
		m+p\cdot \sigma & 0 \\
		0 & m+p\cdot \overline{\sigma} \\
\end{bmatrix}.
\end{align}  
Note that the boost generator $S^{0k}$ is not  Hermitian,
therefore the implementation of the homogeneous 
Lorentz group is not unitary 
\cite{vanhees}.

Now we can calculate the spinor with arbitrary 4-momentum $p=(E,\vec{p})$,
from the spinor in the rest frame, i.e. $u(p,\alpha)=D(\eta)u(p_{0},\alpha)$:
\begin{align}\label{7}
u(p,\alpha)=\frac{N}{\sqrt{2m(E+m)}}
\begin{bmatrix}
		(m+p\cdot \sigma)\xi ^{\alpha} \\
		(m+p\cdot \overline{\sigma})\xi ^{\alpha} \\
\end{bmatrix}.
\end{align} 
Alternatively, we can also write:
\begin{align}\label{8}
u(p,\alpha)=\frac{N}{\sqrt{m}}
\begin{bmatrix}
		\sqrt{p\cdot \sigma}\xi ^{\alpha} \\
		\sqrt{p\cdot \overline{\sigma}}\xi ^{\alpha} \\
\end{bmatrix}.
\end{align} 
Analogously for the spinors  $v(p_{0},\alpha)$ (corresponding to the
antiparticle), we have:
\begin{align}\label{10}
v(p,\alpha)=\frac{M}{\sqrt{m}}
\begin{bmatrix}
		\sqrt{p\cdot \sigma}\eta ^{\alpha} \\
		-\sqrt{p\cdot \overline{\sigma}}\eta ^{\alpha} \\
\end{bmatrix}.
\end{align} 

To calculate the normalization constants we need to define the dual of the
 Dirac spinors.
 A good choice is
\begin{equation}
 \overline{u}(p,\alpha)=u^{\dag}(p,\alpha)\gamma^{0}, \,\,\, 
 \overline{v}(p,\alpha)=v^{\dag}(p,\alpha)\gamma^{0}. 
\end{equation}
The 
justification of such a choice is that it  guarantees the Lorentz
 invariance of the inner product.
 With $\overline{u}(p,\alpha)u(p,\alpha)=1$ and
 $\overline{v}(p,\alpha)v(p,\alpha)=1$, we obtain for the normalization
constants:  $N=1/\sqrt{2}$ and $M=1/\sqrt{-2}$. Finally,  the Dirac spinors 
in an arbitrary frame can be written as:
\begin{align}\label{16}
u(p,\alpha)=\frac{1}{\sqrt{2m}}
\begin{bmatrix}
		\sqrt{p\cdot \sigma}\xi ^{\alpha} \\
		\sqrt{p\cdot \overline{\sigma}}\xi ^{\alpha} \\
\end{bmatrix},
\end{align}
\begin{align}\label{16b}
v(p,\alpha)=\frac{1}{\sqrt{-2m}}
\begin{bmatrix}
		\sqrt{p\cdot \sigma}\eta ^{\alpha} \\
		-\sqrt{p\cdot \overline{\sigma}}\eta ^{\alpha} \\
\end{bmatrix}.
\end{align}
With these expressions we have a complete and orthonormal basis for
 the Hilbert space of the spinors. This space is denoted
 by $\mathcal{H}(m,1/2,\pm)$,  which is a direct sum of the Hilbert
 space of the particle (rest mass $m$, spin $1/2$ and positive energy)
 and the Hilbert space of its antiparticle (rest mass $m$, spin $1/2$,
 but with negative energy) \cite{vanhees}. 
The Hilbert space  labels come from the Casimir operators of the
 Lorentz and Poincar\'e group, 
which are $p^{\mu}p_{\mu}=m^2$ and $S^{2}$.
 Therefore the mass and spin are the kinematic 
labels for the quantum systems \cite{wigner39}.
The attentive reader certainly  noted that the spinors (Eqs.\ref{16} and \ref{16b})
 are not tensor products of kets of momentum and spin 
(e.g. $u(p,\alpha)\neq \ket{p}\ket{\alpha}$), 
and it is not possible to transform  spin 
independently of momentum (Eqs.\ref{15}, \ref{15b} and \ref{7}),
 let alone a transformation on a
 {\em reduced spin  density matrix}, which is ill defined in the 
relativistic context \cite{czachor03}.

\section{ The  spinor density matrix}
In this section we start discussing general superpositions of spinors,
then we introduce mixed states, and investigate covariance
properties of expectation values,  and finally unify the representation 
of pure and mixed states in a Bloch sphere. 
The motivation is to show that a properly defined mixed state should
be covariant, which has the important consequence of the invariance
of pure states' entanglement  under Lorentz transformations. This
conclusion, though obvious, seems to be  still misunderstood in the literature
\cite{contraexemplo}.

It is important to stress that we are treating a free Dirac particle,
whose Hilbert space is a direct sum of the particle's and antiparticle's
sub-spaces, 
 $\mathcal{H}(m,1/2,\pm)=\mathcal{H}(m,1/2,+)\oplus\mathcal{H}(m,1/2,-)$.
Only CPT transformations (i.e., inversion of both charge and parity,  and
 time reversal)  
can connect these sub-spaces  \cite{peskin}, transforming a particle
in an antiparticle, and vice-versa. As CPT transformations are not
considered in the present context, 
 we can write a pure spin state for a particle with
 momentum $p=(E,\vec{p})$ as:
\begin{equation}\label{23}
\psi(p)=\sum_{\alpha=0}^{1}a(\alpha)u(p,\alpha),
\end{equation} 
with $\sum_\alpha|a(\alpha)|^2=1$.
 If  Eq.\ref{23} represents a quantum state, the 
squared coefficients $|a(\alpha)|^2$ must correspond
to some  probability distribution.
We can check if this is so by means of the  conservation of the 4-current:
\begin{equation} 
\partial_{\mu}j^{\mu}=0. 
\end{equation}
 $j^{0}$ is the probability distribution, 
 and $\vec{j}$ is the probability current associated with
 the spinor $\Psi(x)$, such that
 $j^{\mu}=(j^{0},\vec{j})=(\Psi(x)^{\dagger}\Psi(x),\Psi(x)^{\dagger}\vec{\alpha}\Psi(x))$.
$\Psi(x)$ is a solution of the Dirac equation,
\begin{equation}
\Psi(x)=\sum_{\alpha} a(\alpha)u(p,\alpha)\exp{(-ip\cdot x)}.
\end{equation}

Now the probability distribution $j^0$ can be written in the
 basis $\{u(p,\alpha)\}$ as:
\begin{equation} 
 j^0=\sum_{\alpha,\beta} a(\alpha)^{*} a(\beta) u(p,\alpha)^{\dagger}u(p,\beta), 
\end{equation}
and using the inner product relation
 $u(p,\alpha)^{\dagger}u(p,\beta)=\frac{E}{m} \delta_{\alpha,\beta}$, 
we finally arrive at:
\begin{equation} \label{j0}
 j^0=\frac{E}{m} \sum_a|a(\alpha)|^{2}.
\end{equation}
Therefore, as $E/m$ is a positive constant, the coefficients 
$|a(\alpha)|^2$ are the $j^0$ probability distribution coefficients. 

Now we check that though  the Lorentz boost is not unitary,
 the normalization of the spinor is  invariant under Lorentz
 transformations \cite{greiner}. 
Let the dual of the spinor $\psi(p)$ in the frame $S$ be: 
\begin{equation}\label{dual}
\bar{\psi}(p)=\psi(p)^\dag \gamma^0. 
\end{equation}
In  other frame $S'$, the spinor  $\psi'(p')$ and  its dual
$\bar{\psi'}(p')$  are related to   frame $S$ through the
 Lorentz transformation $D(\omega)=\exp{(-i\omega_{\mu\nu}S^{\mu\nu}/2)}$:
\begin{align}\label{21}
\psi'(p')=D(\omega)\psi(p)
\quad \text{and} \quad
\bar{\psi'}(p')=\bar{\psi}(p)D^{-1}(\omega).
\end{align}
The transformation for the spinors is trivial, but it is not
so obvious in the case of the duals, for we have to use the 
inverse Lorentz transformation given by
\begin{equation}\label{lorentzinv}
D^{-1}(\omega)=\gamma^{0}D^{\dagger}(\omega)\gamma^{0}.
\end{equation}
 The explicit calculation is as follows:
\begin{eqnarray}
\bar{\psi'}(p')&=&\psi'^{\dagger}(p')\gamma^{0} \\ 
	       &=&\psi^{\dagger}(p)D(\omega)^{\dagger}\gamma^{0} \\
	       &=&\psi^{\dagger}(p)\gamma^{0}D(\omega)^{-1}\\
               &=&\bar{\psi}(p)D(\omega)^{-1}.
\end{eqnarray}
Finally we have:
\begin{equation}\label{20}
\bar{\psi}'(p')\psi'(p')=\bar{\psi}(p)\psi(p),
\end{equation} 
which implies that the normalization is invariant under
 Lorentz transformations, as expected.

Now that we have well defined pure sates, we introduce the mixed states
by means of the convex sum \cite{strazhev84}:
\begin{equation}\label{24}
\rho(p)=\sum_{k}q_{k}\psi_{k}(p)\overline{\psi}_{k}(p),
\end{equation}  
where $\psi_{k}(p)$ are pure states given by Eq.\ref{23} and 
$q_{k}$ represents the probability to obtain the $k^{th}$ state, 
such that $\sum_{k}q_{k}=1$.

Now we shall calculate how these mixed states behave under
 Lorentz transformations.
  A state $\rho'(p')$ in the  frame $S'$ is written as:
\begin{equation}
\rho'(p')=\sum_{k}q_{k}\psi'_{k}(p')\overline{\psi}'_{k}(p').
\end{equation} 
Invoking the Lorentz transformation Eq.\ref{21},   it is evident that:
\begin{equation}
\rho'(p')=\sum_{k}q_{k}D(\omega)\psi_{k}(p)\overline{\psi}_{k}(p)D(\omega)^{-1},
\end{equation} 
and therefore 
\begin{equation}\label{25}
\rho'(p')=D(\omega)\rho(p)D(\omega)^{-1}.
\end{equation}

Now we check that the expectation value of a Hermitian operator
 ($A(p)$) is the same in all frames:  
\begin{eqnarray*}
Tr[A'(p')\rho'(p')]&=&Tr[D(\omega)A(p)D(\omega)^{-1}D(\omega)\rho(p)D(\omega)^{-1}]\\
Tr[A'(p')\rho'(p')]&=&Tr[A(p)\rho(p)].
\end{eqnarray*}
It follows that the  trace and eigenvalues of the density 
matrix ($\rho$) are covariant.
Of course, to maintain invariance, we must transform both state and observable.

Another important consequence of the covariance of expectation values is 
the invariance of the purity (or mixedness)  of the density matrix, i.e. 
$Tr[\rho'(p')^{2}]=Tr[\rho(p)^{2}].$
A corollary of this straightforward result is the invariance of  entanglement
 of pure states  under Lorentz transformations, for in this case the purity
of the marginals characterize the entanglement.

Let us write a Bloch like decomposition for the density matrix
 \cite{strazhev84}. First we define the Pauli matrices decomposed in the 
Dirac spinors as:
\begin{eqnarray}
\Sigma_{x}(p)&=& u(p,0)\overline{u}(p,1)+u(p,1)\overline{u}(p,0),\\
\Sigma_{y}(p)&=& i[u(p,1)\overline{u}(p,0)-u(p,0)\overline{u}(p,1)],\\
\Sigma_{z}(p)&=&u(p,0)\overline{u}(p,0)-u(p,1)\overline{u}(p,1).
\end{eqnarray}
The matrix $I(p)/2$ is the maximally mixed (or depolarized) state,
 which can be decomposed  in the Dirac spinors as:
\begin{equation} \label{28}
 \frac{I(p)}{2}=\frac{1}{2}\sum_{\alpha}u(p,\alpha)\overline{u}(p,\alpha).
\end{equation}
$I(p)$ also satisfies the relation \cite{peskin}:
\begin{equation}
I(p)=\sum_{\alpha}u(p,\alpha)\overline{u}(p,\alpha)=(\gamma^{\mu}p_{\mu}+m)/2m.
\end{equation}

Finally we have a continuous manifold, for $p$ is continuous, 
that has a Bloch like sphere in each definite momentum $p=(E,\vec{p})$: 
\begin{equation}\label{27}
 \rho(p)=\frac{1}{2}I(p)+\frac{1}{2}r_{l}\Sigma^{l}(p).
\end{equation}
As in the nonrelativistic case, we have  pure states for 
$|\vec{r}|=1$, and mixed states for $|\vec{r}|<1$. 
Note that to each momentum value there is associated  a Bloch sphere on spins, 
such that the spin quantization axis, or the antipodal points on a 
canonical basis orientation (like $\sigma_{z}$ eigenstates), 
depends explicitly on the momentum of the reference frame.      

Note that we have defined a probability distribution Eq.\ref{j0} on 
the spin degrees of freedom, because we assumed a particle with a known
definite momentum. In a quantum information context this is justified
because two parties interested in performing some task, a protocol, 
must share a common reference frame, in order to have a well defined
quantization axis. Remember, for instance, that in the quantum teleport 
protocol \cite{bennett}, Alice must inform Bob the directions she
measured spin, in order to Bob perform his measurements to recover
the quantum information.

\section{Spin quantization axis under Lorentz transformations}

In this section we discuss how a spin measurement depends on the
particle's momentum under Lorentz transformations. The scenario is
a source, in the rest frame $S$ on Earth, emitting particles  with  velocity
$v\hat{z}$ and spin up in  the $z$ direction, such that
\begin{equation}
S_{z}u(m,0)=\frac{1}{2}u(m,0).
\end{equation}
An observer at rest in a frame $S'$ on a satellite,
 which moves with velocity $-\beta\hat{x}$ in relation to $S$, 
measures the spin of the particles.
Our problem is to preview the measurement outcomes in $S'$ as a
 function of the detector's orientation. 

We start by writing the momentum of the particle in the frame $S$,
\begin{equation}\label{a}
p=m(\cosh{\eta},0,0,-\sinh{\eta}).
\end{equation}
This momentum is obtained performing a Lorentz transformation (boost in 
$z$ direction)  on a
particle at rest ($p_{0}=(m,0,0,0)$),
\begin{equation}
p=L(\eta)p_{0},
\end{equation}
where
\begin{align}
L(\eta)=
\begin{bmatrix}
		\cosh{\eta} & 0 & 0 & -\sinh{\eta} \\
		0 & 1 & 0 & 0 \\
		0 & 0 & 1 & 0 \\
		-\sinh{\eta} & 0 & 0 & \cosh{\eta} \\
\end{bmatrix},
\end{align}
being $\eta$  the particle's rapidity ($\tanh{\eta}=v$).

The particle's momentum in the satellite ($p'$) is obtained
by means of a boost in the $-x$ direction on $p$:
\begin{equation}
p'=L(\omega)p,
\end{equation}
where  
\begin{align}
L(\omega)=
\begin{bmatrix}
		\cosh{\omega} & -\sinh{\omega} & 0 & 0 \\
		-\sinh{\omega} & \cosh{\omega} & 0 & 0 \\
		0 & 0 & 1 & 0 \\
		0 & 0 & 0 & 1 \\
\end{bmatrix},
\end{align}
being $\omega$  the rapidity of the satellite ($\tanh{\omega}=\beta$).
 Finally we arrive at $p'=(p'_0,\vec{p'})$, where
\begin{eqnarray}\label{100}
&&p'_0= m\cosh{\omega}\cosh{\eta},\\
		&&\vec{p'}= (m\sinh{\omega}\cosh{\eta},0,-m\sinh{\eta}).
	\end{eqnarray}

Substituting the momentum $p'$ in Eq.\ref{7}, we obtain the particle's
 spinor in the satellite:
\begin{align}
u(p',\alpha)=\frac{1}{\sqrt{4m(p'_0+m)}}
\begin{bmatrix}
		(m+p'_{0}-\vec{p'}\cdot\vec{\sigma})\xi^{\alpha} \\
		(m+p'_{0}+\vec{p'}\cdot\vec{\sigma})\xi^{\alpha}\\
\end{bmatrix},
\end{align}
Performing the matrix products in the above equation  results in: 
\begin{align}
u(p',\alpha)=\frac{1}{\sqrt{4m(p'_{0}+m)}}
\begin{bmatrix}
	(m+p'_{0}-(-1)^{\alpha}p'_{z})\xi^{\alpha}-p'_{x}\xi^{\alpha+1} \\
	(m+p'_{0}+(-1)^{\alpha}p'_{z})\xi^{\alpha}+p'_{x}\xi^{\alpha+1}\\
\end{bmatrix}.
\end{align}

In order to define the spin measurement in the satellite, 
we introduce the following projectors: 
\begin{equation}
P_{\pm}= \ketbra{\theta,\phi,\pm}{\theta,\phi,\pm},
\end{equation}
where
\begin{eqnarray}
\ket{\theta,\phi,+}=\cos\theta/2 \xi^{0}+ \exp{(i\phi)}\sin\theta/2\xi^{1},\\
\ket{\theta,\phi,-}=\sin\theta/2 \xi^{0}- \exp{(i\phi)}\cos\theta/2\xi^{1}.
\end{eqnarray} 
The spin measurement in the satellite reveals that, in relation to Earth,
the quantization axis changes by $\theta$ and the spinor gains a 
relative phase $\phi$.
The spin measurement operator has to belong to both the Lorentz algebra
and the angular momentum Lie algebra. A pair of operators satisfying these
restrictions and defining a complete  measurement is: 
\begin{align}M_{\pm}=
\begin{bmatrix}
 		\ketbra{\theta,\phi,\pm}{\theta,\phi,\pm} & 0 \\
 		0 & \ketbra{\theta,\phi,\pm}{\theta,\phi,\pm}\\
 \end{bmatrix}.
\end{align}

\begin{figure}[t]\label{fig1}	
	
	\includegraphics[scale=0.7]{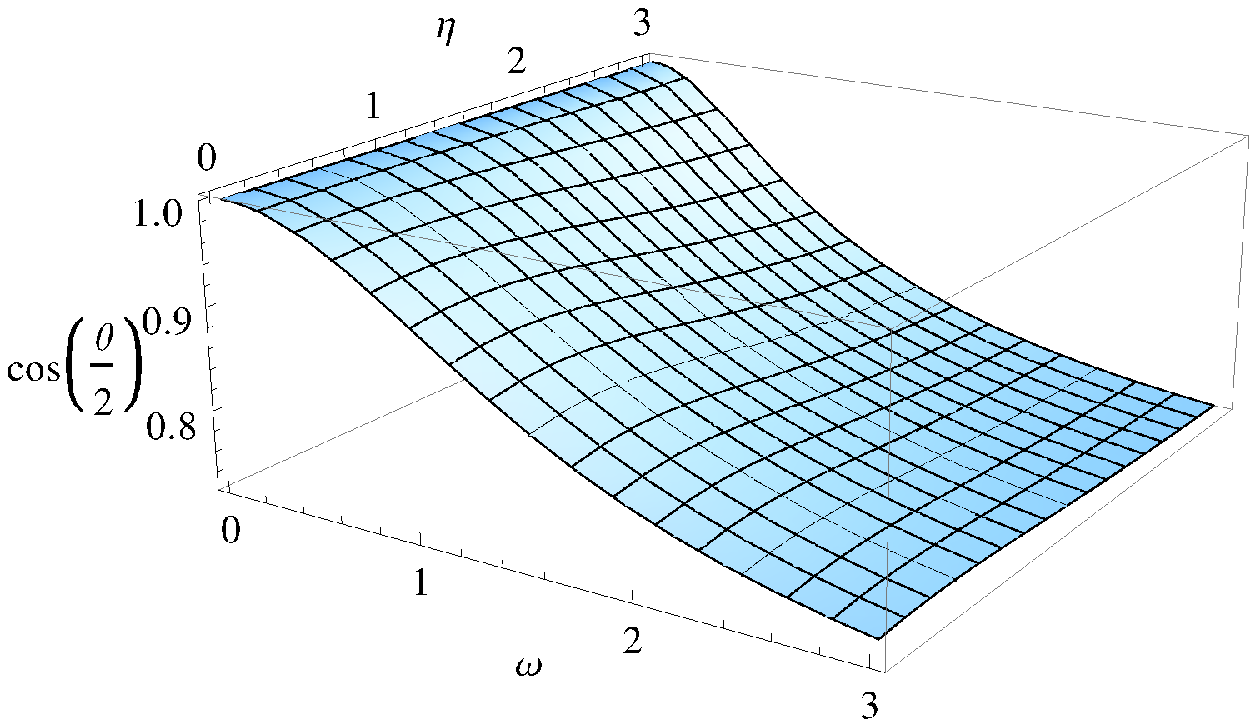} \qquad
	\includegraphics[scale=0.7]{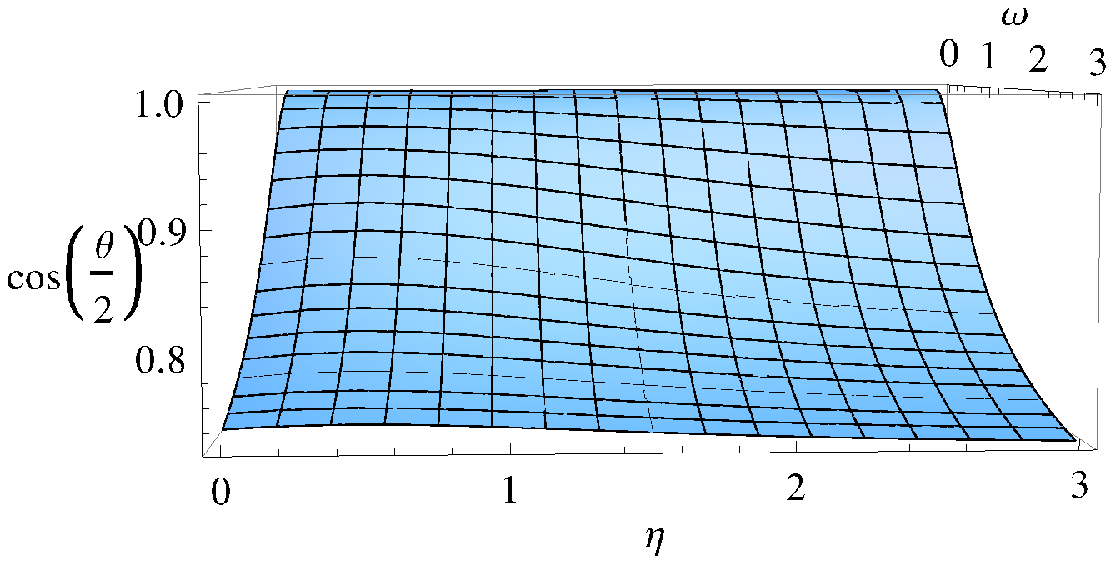}
		\caption{
(First panel) Inclination ($\theta$) of the spin quantization axis  on the
  satellite in relation to Earth, 
as a function of the rapidities of the particle ($\eta$) and of the satellite 
($\omega$).
(Second panel) The effect of the speed of the particle is just to attenuate
how the quantization axis ($z$ in the rest frame) goes to the $x$ direction 
with increasing rapidities.}
\end{figure}

As expectation values are Lorentz invariant, and our particle was 
prepared on Earth in a state with spin up in the $z$ direction, 
if the detectors in the satellite are properly oriented, we 
should obtain:
\begin{equation}
\bar{u}(p',0)M_{+}u(p',0)=1
\end{equation}
and
\begin{equation}
\bar{u}(p',0)M_{-}u(p',0)=0.
\end{equation}
These expectation values lead to the following non-linear system of
equations for $\theta$ and $\phi$:
\begin{align}\label{400}
\bar{u}(p',0)M_{+}u(p',0)=\frac{1}{2m(m+{p'}_{0})}\Big\{[(m+{p'}_{0})^2-{p'}_{z}^{2}]\cos^{2}{\theta/2}-{p'}_{x}^{2}\sin^2{\theta/2}- \\ \nonumber
-2{p'}_{x}{p'}_{z}\cos{\theta/2}\sin{\theta/2}\cos{\phi}  \Big\}=1,\\
\bar{u}(p',0)M_{-}u(p',0)=\frac{1}{2m(m+{p'}_{0})}\Big\{[(m+{p'}_{0})^2-{p'}_{z}^{2}]\sin^{2}{\theta/2}-{p'}^{2}_{x}\cos^2{\theta/2}-\\ \nonumber
-2{p'}_{x}{p'}_{z}\cos{\theta/2}\sin{\theta/2}\cos{\phi}  \Big\}=0.
\end{align}
The solution of the non-linear system results in  null relative 
phase ($\phi=0$), and the angle $\theta$ depends on both the 
rapidities of the particle ($\eta$) and of the satellite ($\omega$),
as expected.
In Fig.1,  we see that
 the quantization axis tends to $\hat{z}$
 ($\cos\theta/2=1$), as the particle's
 velocity tends to zero, as expected.
 On the other hand, for a particle moving near to the speed of light,
the spin quantization axis tends to $\hat{x}$. 
Of course a massive particle never reaches the speed of light, and the
plot in Fig.1 never touches the $x$ axis. This result is nice, for
it is well known that a massless particle always  has its spin,
 or rather its helicity, parallel to the momentum.

It is interesting to note that the Lorentz transformation, though
not unitary, 
 acts on the spin degree of freedom like a rotation in
 the quantization axis. This rotation is like a little group 
representation of the Poincar\'e (inhomogeneous) group, 
and  belongs to $SU(2)$  \cite{ryder98}.
Therefore, as entanglement is invariant under local unitaries, 
we conclude that the {\em entanglement} of a  system 
under Lorentz transformation cannot
change, what changes is just the spin quantization axis as 
a function of  momentum, as we see in Fig.1. 

We could analyze a  simpler situation where the particle is at rest,
with spin  up, and it is to be measured by a moving observer with
rapidity $\omega$. This observer sees the spin quantization axis
rotate according to (Fig.2): 
\begin{equation}
\cos^{2}\theta/2=\frac{2(1+\cosh\omega)+\sinh^2\omega}{(1+\cosh\omega)^2+\sinh^2\omega}.
\end{equation} 
This result is just an evidence that what really matters for the
{\em relativity principle} is the {\em relative movement}.

\begin{figure}[t]
\centering 
\label{fig2} 
\includegraphics[scale=0.70]{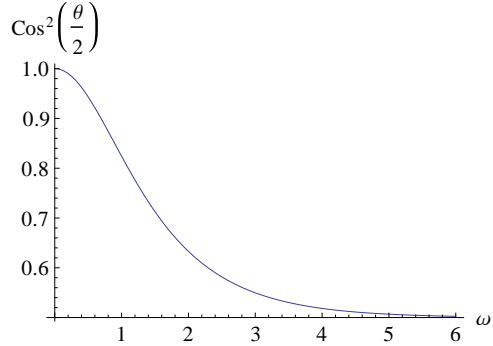} 
\caption{ The inclination ($\theta$) of the spin quantization
axis goes from $0$ 
to $\pi/2$ as the particle's  speed tends to the speed of light.}
\end{figure} 

\section{Conclusion}

In order to discuss quantum information in the relativistic context, 
one has first to properly describe one-particle states under Lorentz
transformations, and that was what we did. 
After revising the Dirac equation for a free particle, we obtained
the covariance of expectation values, which  implies the covariance
of the eigenvalues of the spinor density matrix. 
The covariance of the eigenvalues also  implies  the covariance of 
the  purity of the density matrix. It follows then that entanglement of
a bipartite pure state is covariant,
for it  can be characterized by the purity of
the marginal density matrix. We saw that what changes in a spin-$\frac{1}{2}$
particle under Lorentz transformations is
 the spin quantization axis as a function
of the momentum. The Lorentz transformation acts on the spin degree
of freedom as a local rotation, and this is another way to understand why 
the entanglement  does not change.
As a matter of fact, from the point of view of Lorentz transformations,
spin and momentum are not independent degrees of freedom, but just
labels of the Hilbert space. 
Finally, in Fig.1 and Fig.2, we illustrated how the detectors should be
aligned, depending on the momentum, for a proper 
spin measurement. Note that if the observer in the moving frame ignores
his momentum in relation to Earth, and therefore cannot calculate the
proper alignment of his detectors according to the Lorentz transformation,
this is just {\em classical ignorance} and cannot induce any quantum 
effect.

{\em Acknowledgments} - We thank M.C. Nemes, M.D.R. Sampaio
for the discussions. To J. C. Brant by her intellectual support. 
Financial support by the
Brazilian agencies  FAPEMIG, CNPQ and  INCT-IQ (National 
Institute of Science and Technology for Quantum Information)


\end{document}